\documentclass[twocolumn]{aastex631}

\usepackage{physics,bm, enumerate}

\begin{document}

\title{
Semi-analytical model for the dynamical evolution of planetary systems via giant impacts
}

\author[0000-0001-8477-2523]{Tadahiro Kimura}
\email{tad.kimura624@gmail.com}
\affiliation{UTokyo Organization for Planetary Space Science (UTOPS), University of Tokyo, Hongo, Bunkyo-ku, Tokyo 113-0033, Japan}
\affiliation{Kapteyn Astronomical Institute, University of Groningen Landleven 12, 9747 AD, Groningen, Netherlands}
\affiliation{Division of Science, National Astronomical Observatory of Japan,
Osawa, Mitaka, Tokyo 181-8588, Japan}

\author[0000-0001-5953-3897]{Haruka Hoshino}
\affiliation{Division of Science, National Astronomical Observatory of Japan,
Osawa, Mitaka, Tokyo 181-8588, Japan}
\affiliation{Department of Astronomy, University of Tokyo,
Hongo, Bunkyo-ku, Tokyo 113-0033, Japan}

\author[0000-0002-5486-7828]{Eiichiro Kokubo}
\affiliation{Division of Science, National Astronomical Observatory of Japan,
Osawa, Mitaka, Tokyo 181-8588, Japan}
\affiliation{Center for Computational Astrophysics, National Astronomical Observatory of Japan, 
Osawa, Mitaka, Tokyo 181-8588, Japan}
\affiliation{Department of Astronomy, University of Tokyo,
Hongo, Bunkyo-ku, Tokyo 113-0033, Japan}

\author[0000-0002-2383-1216]{Yuji Matsumoto}
\affiliation{Center for Computational Astrophysics, National Astronomical Observatory of Japan, 
Osawa, Mitaka, Tokyo 181-8588, Japan}

\author[0000-0002-5658-5971]{Masahiro Ikoma}
\affiliation{Division of Science, National Astronomical Observatory of Japan,
Osawa, Mitaka, Tokyo 181-8588, Japan}
\affiliation{Department of Earth and Planetary Science, University of Tokyo, Hongo, Bunkyo-ku, Tokyo 113-0033, Japan}



\begin{abstract}
In the standard model of terrestrial planet formation, planets are formed through giant impacts of planetary embryos after the dispersal of the protoplanetary gas disc.
Traditionally, $N$-body simulations have been used to investigate this process.
However, they are computationally too expensive to generate sufficient planetary populations for statistical comparisons with observational data.
A previous study introduced a semi-analytical model that incorporates the orbital and accretionary evolution of planets due to giant impacts and gravitational scattering. 
This model succeeded in reproducing the statistical features of planets in $N$-body simulations near 1 au around solar-mass stars.
However, this model is not applicable to close-in regions (around 0.1 au) or low-mass stars because the dynamical evolution of planetary systems depends on the orbital radius and stellar mass.
This study presents a new semi-analytical model applicable to close-in orbits around stars of various masses, validated through comparison with $N$-body simulations.
The model accurately predicts the final distributions of planetary mass, semi-major axis, and eccentricity for the wide ranges of orbital radius, initial planetary mass, and stellar mass, with significantly reduced computation time compared to $N$-body simulations.
By integrating this model with other planet-forming processes, a computationally low-cost planetary population synthesis model can be developed. 
\end{abstract}



\section{Introduction} \label{sec:intro}
Exoplanet exploration has revealed many 
planets with radii of 1--4$R_\oplus$, known as super-Earths or sub-Neptunes, exist at short orbital periods $<$~100~days. Nearly half of FGK stars are estimated to have at least one such planet, and many of them harbour multiple planets~\citep{Howard2013,Zhu+2018}. The abundance of such small-sized planets has enabled population studies to, for instance, identify a distinct gap between the super-Earth and sub-Neptune populations, known as the radius valley~\citep[e.g.,][]{Fulton+2017,Fulton+2018,Martinez+2019}. Theories have struggled to explain the origins of the statistical properties of these small planets. 

These planets are thought to form through efficient inward orbital migration, gathering near the central star~\citep[e.g.,][]{Papaloizou+Larwood2000,Brunini+Cionco2005,Terquem+Papaloizou2007,Ida+Lin2010,Coleman+Nelson2014,Cossou+2014, Huang+Ormel2022}. 
After the protoplanetary disc disperses, the planets undergo dynamical evolution such as orbital crossing and collisions, determining their final masses and orbits~\citep[e.g.,][]{Ogihara+Ida2009,McNeil+Nelson2010,Izidoro+2017}.
Most of these previous studies performed direct $N$-body simulations to follow the dynamical evolution of multiple planets.
$N$-body simulations are a crucial method for understanding the physical processes involved in planet formation. However, their computational cost is quite high, making it challenging to perform a sufficient number of simulations for statistical comparison with observations.

The population synthesis method has been proposed to make statistical comparisons between theoretical results and observations~\citep[e.g.,][]{Ida+Lin2004,Mordasini+2009}.
Recent models adopting $N$-body integrations require nearly one million CPU hours to synthesize a number of planets comparable to those currently detected in exoplanetary systems~\citep{Emsenhuber+2021a,Emsenhuber+2021b}.
To reduce the computational cost, \cite{Ida+Lin2010} developed a semi-analytical model of this dynamical evolution
by using a Monte Carlo approach to determine the changes in orbital elements due to collisions and close scatterings.
Although, in general, the orbital evolution of each planet due to the gravitational interaction with many bodies is highly chaotic, the final planetary mass and orbital distribution show some statistical features~\citep[e.g.,][]{Chambers2001, Kokubo+2006,Matsumoto+2015,Hoffmann+2017,Goldberg+Batygin2022}.
\cite{Ida+Lin2010} showed that their semi-analytical model could reproduce the results of direct $N$-body simulations for the dynamical evolution of multi-planet systems, especially in terms of the typical masses, semi-major axes, and eccentricities of relatively massive planets in each planetary system.

However, their comparison between analytical simulations and $N$-body simulations is limited to the case of systems of planets orbiting near 1~au around $1M_\odot$ stars, suggesting that the semi-analytical model is not applicable to other regions around Sun-like stars or to other types of stars, where the orbital evolution features and the final planetary mass and orbital distributions differ~\citep[][]{Matsumoto+Kokubo2017,Matsumoto+2020, Hoshino+Kokubo2023}.
In particular, given that most detected exoplanets are located in orbits with periods of $\lesssim 100$~days~\citep[e.g.,][]{Howard+2010,Zhu+2018,Weiss+2023}, it is crucial to develop a new model applicable to those in such close-in regions.

Here, we develop a new semi-analytical dynamical evolution model that can reproduce various statistical results from $N$-body simulations for wide ranges of orbital radii and stellar masses.
In \S~\ref{sec:method}, we present the details of the semi-analytical model. The validation of the model through the comparison with the $N$-body simulations is presented in \S~\ref{sec:results}.
Then we discuss some limitations of our model and future perspectives in \S~\ref{sec:discussion}, and
finally, in \S~\ref{sec:conclusion}, we summarise our results and discuss future perspectives.

\section{semi-analytical model} \label{sec:method}

We describe the semi-analytical model for the dynamical evolution of multiple planets.
We should note that in this paper, ``planet'' refers to relatively low-mass planets such as terrestrial planets and super-Earths/sub-Neptunes, and does not include gas giants.
Hereafter we denote the mass, semi-major axis, eccentricity, inclination, and orbital period of the planet $i$ as $M_i, a_i, e_i$, $I_i$, and $P_i$, respectively.
The semi-analytical model includes the following processes;
\begin{enumerate}[(1)]
    \item Evolution of planet eccentricity via secular perturbations 
    \item Determination of when and which planetary pair will have an orbital crossing event
    \item Calculation of the probability that the planetary pair undergoing orbital crossing eventually collide with each other
    \item Calculation of the new orbital elements after the orbital crossing event
\end{enumerate}
We should note that the inclination of the planets is assumed to be half of the eccentricity in the entire simulations.
The details of each process are provided below.
See Appendix~\ref{sec:semi_analytical_model} for the calculation procedure for the entire model combining the above processes.

\subsection{Eccentricity evolution by secular perturbation}

Orbital eccentricity is a crucial factor in determining how the planetary system dynamically evolves.
In our model, we analytically calculate the evolution of the eccentricity due to secular perturbations from other planets~\cite[e.g.,][]{Murray+Dermott1999}.
Assuming that $e_i \ll 1$ and $I_i=0$, the secular evolution of the eccentricity vector of planet $i$ in an $N$-body system can be analytically expressed by
\begin{align}
    h_i(t) &:= e_i \sin \varpi_i = \sum_{j=1}^N e_{ij}\sin (g_j t + \beta_j), 
    \label{eq:ecc_vector_h}
    \\
    k_i(t) &:= e_i \cos \varpi_i = \sum_{j=1}^N e_{ij}\cos (g_j t + \beta_j),
    \label{eq:ecc_vector_k}
\end{align}
with $\varpi_i$ being the longitude of pericenter. 
See Appendix~\ref{sec:secular_theory} for derivation of the coefficients and angles $e_{ij}, g_j$, and $\beta_j$.  
Then the eccentricity is simply calculated by
\begin{equation}
    e_i(t) = \sqrt{h_i(t)^2 + k_i(t)^2}.
    \label{eq:ecc_sec}
\end{equation}

\subsection{Timescale until the onset of orbital instability}

To calculate the dynamical evolution of a planetary system, it is crucial to predict when the system becomes dynamically unstable, leading to events such as collisions or close scattering.
In our semi-analytical model, we calculate the time until each adjacent pair of planets experiences an orbital crossing, referred to as the orbital crossing timescale $\tau_{\rm cross}$. Here we use the analytical formula given by \cite{Petit+2020},
which is based on the analysis of the diffusion timescale of planetary orbits via three-body resonances.
See Appendix~\ref{sec:tcross_by_Petit} for the explicit form of the formula.

Since, however, their formula is for a system with three circular-orbit planets, we have made a few modifications to apply the formula to eccentric $N$-body systems.
Firstly, $\tau_{\rm cross}$ depends significantly on
the normalized orbital separation of adjacent planets ($i,j; a_i < a_j$), which is defined as $\delta_{ij} = (a_j-a_i)/a_j$ in \cite{Petit+2020}.
Here we redefine $\delta_{ij}$ by considering the distance between the apocenter of the inner planet and the pericenter of the outer planet, namely the closest distance between the orbits, as follows;
\begin{equation}
    \delta_{ij} = \frac{(1-e_j)a_j - (1+e_i)a_i}{a_j}
    \label{eq:delta_ij}
\end{equation}
Although this treatment is tentative, previous studies have suggested that replacing the orbital separation with the closest distance provides a reasonable approximation for eccentric systems~\citep[e.g.,][]{Pu+Wu2015}.

Furthermore, to apply $\tau_{\rm cross}$ to general $N$-body systems, we introduce a numerical factor $K$ to the density of three-body resonances (see Appendix~\ref{sec:tcross_by_Petit} for details).
This factor accounts for the assumption that the resonance density in $N$-body systems is $K$ times larger than in the three-planet case.
\cite{Petit+2020} suggested that $K=2$ would be a reasonable estimate for a system with five planets, assuming that both the inner and outer neighbours of the triplet increase the resonance density by 50\% each.
Following their discussions, we calculate $K$ by
\begin{equation}
    K = {\rm min}[0.5(N-3) + 1, 3].
\end{equation}
This assumption allows for up to two neighboring planets inside and outside the triplet to affect the number of resonances.
It is consistent with the feature that $\tau_{\rm cross}$ shows little dependence on $N$ when $N\gtrsim 10$~\citep[e.g.,][]{Chambers+1996,Funk+2010}.
We have found that changing the maximum number of $K$ by unity does not significantly affect the results.

\subsection{Collision probability of the orbit-crossing pair}
Hereafter, we consider the situation in which a neighboring planet pair $(i,j)$ causes orbital crossings. 
The eccentricities of the two planets are no longer determined by Eq.~\eqref{eq:ecc_sec} and are excited so that their epicycle amplitudes overlap.
Assuming the energy equipartition~\citep{Nakazawa+Ida1988,Nakazawa+1989}, the eccentricities at the onset of orbital crossing can be expressed by 
\begin{equation}
    e_i \simeq e_{{\rm cross},i}
    = \frac{\sqrt{M_j} b_{ij}}{\sqrt{M_j}a_i+\sqrt{M}_ia_j},
    \label{eq:e_cross}        
\end{equation}
and similarly for $e_j$.
Then, the planets undergo multiple close encounters and eventually either collide or experience a significant change in their orbits due to close scattering (these are collectively referred to as ``events'' hereafter).
In our model, we evaluate the collision probability from the comparison between the orbital scattering timescale $\tau_{\rm scat}$ and the collision timescale $\tau_{\rm col}$, each of which is defined as follows.

The orbital scattering timescale is defined as the period during which their orbits are significantly changed through repeated close encounters and is equal to the two-body relaxation timescale, which is given by~\citep{Ida1990,Kokubo+Ida2012}
\begin{equation}
    \tau_{{\rm scat},ij} = \frac{1}{n\pi R_g^2v_{\rm ran}\ln \Lambda},
    \label{eq:tau_vis_org}
\end{equation}
where $n$ is the number density of planets, $R_g = G(M_i+M_j)/v_{\rm ran}^2$ is the gravitational radius, $v_{\rm ran}$ is the relative (random) velocity, and $\ln \Lambda$ is the Coulomb logarithm coming from the effects of distant encounters.
The number density $n$ is calculated by
\begin{equation}
    n = \frac{1}{2\pi a_{ij}\times 2I_{ij}a_{ij}\times b_{ij}}
    = \frac{1}{2\pi e_{ij}a_{ij}^2b_{ij}},
   \label{eq:number_density}
\end{equation}
with $e_{ij}=\sqrt{e_i^2+e_j^2}$ and $I_{ij}=\sqrt{I_i^2+I_j^2}$ being the relative eccentricity and inclination, respectively, and $a_{ij} = (a_i+a_j)/2$.
Here we have assumed that $b_{ij} \ll a_{ij}$, $I_{ij} \ll 1$, and $e_{ij} = 2 I_{ij}$.
Then, assuming $v_{\rm ran}=e_{ij}v_{\rm K}$,
$\tau_{{\rm scat},ij}$ is written as
\begin{equation}
    \tau_{{\rm scat},ij} = \frac{4b_{ij}a_{ij}}{\pi (R_i+R_j)^2\ln \Lambda}\qty(\frac{e_{ij}}{e_{\rm esc}})^4 T_{\rm K},
    \label{eq:tau_scat}
\end{equation}
where $R_i$ and $R_j$ are the radii of planets $i$ and$j$, respectively, and $e_{\rm esc}$ is the escape eccentricity defined as \citep[e.g.,][]{Safronov1969, Kokubo+Ida2002}
\begin{equation}
    e_{\rm esc} = \frac{v_{\rm esc}}{v_{\rm K}}
= \frac{\sqrt{2G(M_i+M_j)/(R_i+R_j)}}{\sqrt{GM_*/a_{ij}}}.
\label{eq:e_esc}
\end{equation}

On the other hand, the collision timescale is given by~\citep[e.g.,][]{Safronov1969,Ida+Nakazawa1989,Greenzweig+Lissauer1990}
\begin{align}
    \tau_{{\rm col},ij} &= \frac{1}{n\pi (R_i+R_j)^2 (1+v_{\rm esc}^2/v_{\rm ran}^2)v_{\rm ran}} \notag \\
    &= \frac{b_{ij}a_{ij}}{\pi (R_i+R_j)^2}\frac{1}{1+e_{\rm esc}^2/e_{ij}^2}T_{\rm K}.
    \label{eq:tau_col}
\end{align}
Then, the expected number of ``collision chance'' before the orbits are largely changed due to close scattering can be expressed by 
\begin{equation}
    \lambda = \frac{\tau_{{\rm scat},ij}}{\tau_{{\rm col},ij}}
    = 
    \frac{1}{\ln \Lambda}\qty(\frac{2e_{ij}}{e_{\rm esc}})^2 \qty(1+\frac{e_{ij}^2} {e_{\rm esc}^2}).
    \label{eq:lambda}
\end{equation}
We set $\ln \Lambda = 3$~\citep[e.g.,][]{Ida1990,Kokubo+Ida2000}.

Now, in general, the probability that an event expected to occur $\lambda$ times in a given period will actually occur $k$ times is expressed using a Poisson distribution;
\begin{equation}
    p(k;\lambda) = \frac{\lambda^k}{k!}e^{-\lambda}.
\end{equation}
Then, the collision probability, which means ``the probability of a giant collision event actually occurring at least once under the expected value of $\lambda$'' is
\begin{equation}
    p_{\rm col} = 1-p(k=0;\lambda) = 1-\exp(-\lambda).
    \label{eq:pcol}
\end{equation}
We use this $p_{\rm col}$ to determine whether the planetary pair will undergo a giant collision or orbital scattering.

\subsection{Changes in orbits through an event}

After the onset of orbital crossing, the relative eccentricity $e_{ij}$ is excited to $\sim e_{\rm esc}$ and its distribution is relaxed to a Rayleigh distribution through repeated close encounters~\citep{Ida+Makino1992}.
Thus, $e_{ij}$ just before an event is chosen from the Rayleigh distribution with $\langle e_{ij}^2\rangle^{1/2}=e_{\rm esc}$, and the eccentricities for each planet, $(e_{i0},e_{j0})$, are calculated by assuming energy equipartition.
Then, the orbital elements after the event are calculated depending on whether collision or scattering has occurred, as described below.

\subsubsection{Collision Case}
In the case of a giant collision, we assume that the collision is cohesive and results in perfect merging.
Thus, the mass and semi-major axis of the merged planet are calculated so that the mass and the center of mass are conserved;
\begin{align}
    M_{\rm new} &= M_i + M_j, \label{eq:Mnew} \\
    a_{\rm new} &= \frac{M_ia_i+M_ja_j}{M_{\rm new}}.
    \label{eq:anew}
\end{align}
Here, we denote the variables of the merged planet with subscript ``new''.
The eccentricity of the merged planet is calculated by 
the conservation of the Laplace-Runge-Lenz vector \citep{Matsumoto+2015};
\begin{align}
    M_{\rm new}^2e_{\rm new}^2
    &= M_i^2e_{i0}^2+M_j^2e_{j0}^2 \notag \\
    & \quad +2M_iM_je_{i0}e_{j0}\cos(\Delta \varpi_{ij}).
    \label{eq:Mnew_enew}
\end{align}
Here $\Delta \varpi_{ij} = \varpi_i - \varpi_j$ is the difference in the longitude of the pericenter, which is randomly chosen from a certain range so that the two elliptic orbits have intersections.
See Appendix~\ref{sec:semi_analytical_model} for the detailed method.

\subsubsection{Scattering event}
In the orbital scattering case, 
the eccentricities $(e_i,e_j)$ after the event is set to ($e_{i0}, e_{j0})$,
and the change in orbital separation is assumed to be equal to the sum of the excited epicycle amplitude;
\begin{equation}
    \Delta b_{ij} = e_{i0}a_{i0} + e_{j0} a_{j0}, \label{eq:Dbij}
\end{equation}
where $a_{i0},a_{j0}$ are the semi-major axes of the planets $i,j$ before the scattering event.
Then this change in separation is distributed to each planet so that the center of mass is conserved and the inner and outer planets are, respectively, scattered inwards and outwards, that is,
\begin{align}
    a_i &= a_{i0} -\frac{M_j}{M_i+M_j}\Delta b_{ij}, \label{eq:a_i_after_scat}\\
    a_j &= a_{j0} +\frac{M_i}{M_i+M_j}\Delta b_{ij}.
    \label{eq:a_j_after_scat}
\end{align}
Note that these prescriptions do not explicitly preserve orbital energy and angular momentum. However, we have confirmed that within the parameter sets explored in this study, both quantities are approximately conserved within a few percent, even when frequent scattering events occur.

\begin{figure}
    \centering
    \includegraphics[width=\linewidth]{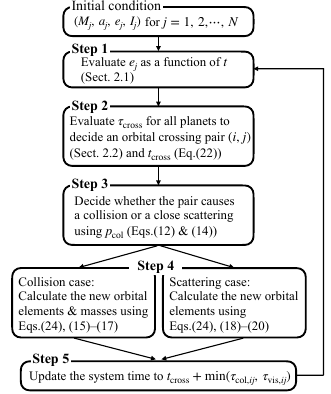}
    \caption{Flowchart of the semi-analytical model. See \S~\ref{sec:semi_analytical_model} for the detailed procedures.}
    \label{fig:flow_chart}
\end{figure}
\subsection{Numerical procedures of semi-analytical model}
\label{sec:semi_analytical_model}

Here we summarize the numerical procedure to calculate the dynamical evolution of planets.
The overview of the model workflow is shown in Fig.~\ref{fig:flow_chart}.
\begin{enumerate}
\item 
    \textbf{Evaluation of eccentricity in each timestep} \\
    From the system architectures (i.e., masses, semi-major axes, and eccentricities of planets) in the previous timestep $t_0$,
    we first evaluate the coefficients $e_{ij}, g_j, \beta_j$ in Eqs.~\eqref{eq:ecc_vector_h} and \eqref{eq:ecc_vector_k} (see Appendix~\ref{sec:secular_theory} for details).
    Since, in our model, the semi-major axes and planetary masses change only when the collision or close scattering event occurs, these coefficients do not change until the next timing of these events comes.
    Thus, the eccentricity of each planet until then can be simply calculated by \eqref{eq:ecc_sec}.
\item \textbf{Timing of the orbital crossing event} \\
    Then we calculate the orbital crossing timescale $\tau_{\rm cross}$ for all three adjacent planets (Eq.~\eqref{eq:tau_cross_Petit}).
    We should note here that simply using the eccentricities at $t_0$ for the calculation of $\tau_{\rm cross}$ is not sufficient, since the eccentricity changes with time via secular evolution.    
    Instead, we use the time-averaged eccentricity, which is evaluated as follows.
    From Eq.~\eqref{eq:ecc_sec}, the average of $e_i^2(t)$ over secular cycle, denoted as $\langle e_i^2(t_0) \rangle$, can be approximated to
    \begin{equation}
        \langle e^2_i(t_0)\rangle \simeq \sum_{j=1}^N e^2_{ij}.
    \end{equation}
    Then we use $\langle e^2_i(t_0)\rangle^{1/2}$ to calculate $\delta_{ij}$ in Eq.~\eqref{eq:delta_ij} and $\tau_{\rm cross}$.
    This is valid when the typical value of $\tau_{\rm cross}$ is much longer than the secular cycle timescale. 
    This is found to be a good approximation in most cases; the eccentricity evolves on a shorter timescale than $\tau_{\rm cross}$ by a few orders of magnitude.
    Thus, if the eccentricity of some planets in a triplet significantly increases (or decreases) through secular evolution, $\tau_{\rm cross}$ becomes shorter (or longer) than the one estimated at $t_0$.\\
    Then, of the planets in the triplet with the shortest $\tau_{\rm cross}$, the two planets $(i,j)$ with the smallest orbital separation actually undergo an orbital crossing at 
    \begin{equation}
        t_{\rm cross} = t_0+\tau^*_{\rm cross}, 
    \end{equation}
    where $\tau^*_{\rm cross}$ is the minimum value of $\tau_{\rm cross}$.
    Hereafter, we assume that $a_i < a_j$.
\item \textbf{Collision probability of the orbit-crossing pair} \\
    Whether the planet pair $(i,j)$ undergoes a giant collision or close scattering is determined according to $p_{\rm col}$ in Eq.~\eqref{eq:pcol}.
    Since the epicycle amplitudes of the two planets must be overlapping at this stage, the eccentricity used for the evaluation of $\lambda$ (Eq.~\eqref{eq:lambda}) is calculated by
    \begin{equation}
        e_i = \max(e_{{\rm cross},i}, e_i(t_{\rm cross}))
        \label{eq:e_during_cross}
    \end{equation}
    with $e_{{\rm cross},i}$ from Eq.~\eqref{eq:e_cross} and $e_i(t_{\rm cross})$ from Eq.~\eqref{eq:ecc_sec}.
\item \textbf{Orbital elements after the event}\\
    Once it has been determined whether collision or close scattering will occur, we calculate the eccentricities just before the event, $(e_{i0}, e_{j0})$.
    The relative eccentricity $e_{ij}$ is first randomly chosen from the Rayleigh distribution with $\langle e^2_{ij}\rangle^{1/2}=e_{\rm esc}$.
    Then, assuming energy equipartition and further ensuring that the epicycle amplitudes of the planets $(i,j)$ overlap, $e_{i0}$ and $e_{j0}$ are calculated as
    \begin{equation}
        e_{i0} = \max\qty(
        \frac{\sqrt{M_j} e_{ij}}{\sqrt{M_i+M_j}},
        e_{{\rm cross},i}, e_i(t_{\rm cross})
        ),
        \label{eq:e_i_before_col}
    \end{equation}
    and similarly for $e_{j0}$. 
    Then the new orbital elements are determined using Eqs.~\eqref{eq:Mnew}--\eqref{eq:Mnew_enew} in the case of collision, and using Eqs.~\eqref{eq:Dbij}--\eqref{eq:a_j_after_scat} in the case of close scattering.
    In the collision case, so that the two elliptical orbits have intersections, the value of $\Delta \varpi_{ij}$ in Eq.~\eqref{eq:Mnew_enew} is determined as follows.
    The distance from the central star to the orbit of the planet $i$ is expressed by
    \begin{equation}
        r_i(\theta_i) = \frac{a_i(1-e_{i0}^2)}{1+e_{i0} \cos (\theta_i-\varpi_i)},
        \label{eq:rj}
    \end{equation}
    with $\theta_i$ being the longitude measured from the reference direction.
    If the two orbits have intersections, $r_i(\theta) > r_j(\theta)$ needs to be satisfied at some range of $\theta$.
    By using Eq.~\eqref{eq:rj}, this can be reduced to the condition:
    \begin{equation}
    \cos \Delta \varpi_{ij} < \frac{e_{i0}^2a_i^2+e_{j0}^2a_j^2 - (a_i-a_j)^2}{2e_{i0}e_{j0} a_i a_j}.
    \end{equation}
    Thus, $\Delta \varpi_{ij}$ is randomly chosen from the range $[\Delta \varpi_{\rm min}, 2\pi-\Delta \varpi_{\rm min}]$, with
    \begin{equation}
        \Delta \varpi_{\rm min} = \acos \frac{e_{i0}^2a_i^2+e_{j0}^2a_j^2 - (a_i-a_j)^2}{2e_{i0}e_{j0} a_i a_j}.
    \end{equation}    
\item \textbf{Update system time} \\
    After performing all the above procedures, the system time is updated to $t_{\rm cross}+\min(\tau_{{\rm scat},ij}, \tau_{{\rm col},ij})$.
\end{enumerate}
These procedures are repeated until the given integration time is reached or the number of planets in the system is reduced to two.
When only two planets are in the system, the orbital crossing timescale 
is no longer valid, and, instead, the stability of the system is determined by the Jacobi energy in the Hill coordinate with the center of mass of the two bodies as the origin~\citep{Nakazawa+Ida1988};
\begin{equation}
    \tilde{E}_J \simeq \frac{1}{2}(\tilde{e}_{ij}^2+\tilde{I}_{ij}^2)-\frac{3}{8}\tilde{b}_{ij}^2 +\frac{9}{2},
    \label{eq:Jacobi}
\end{equation}
where $\tilde{e}_{ij} = e_{ij}/h_{ij}$, $\tilde{I}_{ij} = I_{ij}/h_{ij}$, and $\tilde{b}_{ij} = b_{ij}/r_{\rm H}$, with $h_{ij} = r_{\rm H}/a_{ij}$.
When $\tilde{E}_{\rm J} > 0$, the two planets cross the orbits, and the procedures from Step 3 to 5 are conducted; otherwise, we stop the calculation considering that the system becomes stable.

\begin{table*}
    \centering
    \caption{Initial settings of planetary systems. $M_*$: central stellar mass,
    $\Sigma_0$: initial solid surface density at 1~au,
    $\alpha$: power-law index of initial solid surface density,
    $b_{\rm H,ini}$: initial orbital separation scaled by the mutual Hill radius,
    $a_M$: initial center of mass,
    $N_{\rm ini}$: initial number of planets, $M_{\rm tot}$: total mass of the planets. The values in blank cells are the same as those in the standard model S0.
    }    
    \label{tab:init_cond}
    \begin{tabular}{ccccccccc}\hline
        Model name & orbital radii [au] & $b_{\rm H,ini}$ & $\Sigma_0~[{\rm g~cm^{-2}}]$ & $\alpha$ &  
        $M_* [M_\odot]$ &   $a_M$~[au] & $N_{\rm ini}$ & $M_{\rm tot} [M_\oplus]$ \\ \hline
        S0 & 0.1--0.3 & 10 & 10 & 2.0 & 1 &  0.175  & 15 & 2.43 \\ \hline
        R1 & 0.05--0.15  & & & & &  0.0875 & 15 & 2.43 \\ 
        R2 & 0.2--0.6    & & & & &  0.350  & 15 & 2.43 \\
        R3 & 0.5--1.5    & & & & &  0.875  & 15 & 2.43 \\ \hline
        B1 & & 6  & & & & 0.181  & 34 & 2.55 \\ 
        B2 & & 8  & & & & 0.180  & 22 & 2.54 \\
        B3 & & 12 & & & & 0.181  & 12 & 2.55 \\ \hline
        M1 & & & 5  & & & 0.179  & 22 & 1.26 \\ 
        M2 & & & 20 & & & 0.179  & 11 & 5.03 \\
        M3 & & & 50 & & & 0.180  & 7  & 12.68 \\ \hline        
        A1 & & & & 1.0 & & 0.195  & 38 & 0.45 \\ 
        A2 & & & & 1.5 & & 0.185  & 24 & 1.03 \\ \hline
        S1 & & & & & 0.2 & 0.180  & 7  & 12.68 \\ 
        S2 & & & & & 0.5 & 0.179  & 11 & 5.03 \\
        \hline
    \end{tabular}
\end{table*}

\section{Comparison with {\em N}-body simulation results}
 \label{sec:results}

To validate our semi-analytical model, we compare its results with those of direct $N$-body simulations regarding the characteristics of multi-planetary systems after the dynamical evolution with giant collisions.
The physical significance of the {\em N}-body simulation results will be discussed in a separate paper (Kokubo et al. submitted).

\subsection{Numerical model}

Here, we briefly describe the model and method of the {\em N}-body simulation.
We consider multiple ($N_\mathrm{init}$) protoplanets with the isolation masses in a gas-free environment as initial conditions.
For the calculation of the isolation masses, we assume the solid surface density distribution of
\begin{equation}
    \Sigma = \Sigma_0 \qty(\frac{r}{1~{\rm au}})^{-\alpha}.
\end{equation}
Given that the initial orbital separation of adjacent protoplanets scaled by mutual Hill radius is given by $b_{\rm H,ini}$, the isolation mass is given by~\citep{Kokubo+Ida2002}
\begin{align}
    M_{\rm iso} &= 0.16 \, \qty(\frac{b_{\rm H,ini}}{10})^{3/2} 
    \qty(\frac{\Sigma_0}{10~{\rm g~cm^{-2}}})^{3/2} \notag \\
    & \qquad \times \qty(\frac{a}{1~{\rm au}})^{3(2-\alpha)/2}
    \qty(\frac{M_*}{M_\odot})^{-1/2}
    ~M_\oplus.
\end{align}
We vary the initial range of the orbital radii of protoplanets, as well as the parameters $\Sigma_0$, $\alpha$, $b_{\rm H,ini}$, and $M_*$.
The initial conditions of protoplanet systems are summarised in Table~\ref{tab:init_cond}.
In all models, the bulk density of $3{\rm g~cm^{-3}}$ is assumed when calculating the planet radius.
The initial eccentricities and inclinations are given by the Rayleigh distribution with dispersions of 
\begin{equation}
\langle e^2 \rangle^{1/2} = 2\langle i^2 \rangle^{1/2} = 0.01\qty(\frac{\Sigma_0}{10~{\rm g~cm^{-2}}})^{1/2}.
\end{equation}
For each model, 20 runs are performed with different initial angular distributions of protoplanets for $N$-body simulations and with different random numbers for the semi-analytical model.
The integration time is $5\times 10^8$ Kepler periods of the innermost planet.
The orbital integrations are performed using the modified Hermite scheme~\citep{Kokubo+Ida1998,Kokubo+Makino2004} with the hierarchical time step~\citep{Makino1991}.
All collisional events are assumed to result in perfect accretion.

\subsection{Statistical features to be compared}

When comparing the results of the semi-analytical model and the $N$-body simulations, we focus on the following statistical features of the final distribution of planets.
In this paper, $\bar{m}$ means the averaged value of the quantity $m$ in each run, and $\langle m \rangle$ means the averaged value over 20 runs.

\begin{enumerate}   
    \item \textbf{Final number of planets}\\
    The final number of planets averaged over 20 runs, $\langle N_{\rm p}\rangle$.
    \item \textbf{Mean orbital separation and eccentricity}\\
    The mean orbital separation and eccentricity of adjacent planets, both normalized by the mutual Hill radius, are calculated by
    \begin{align}
        \bar{b}_{\rm H} &= \frac{1}{N_{\rm p}-1}\sum_{i}^{N_{\rm p}-1} \frac{a_{i+1}-a_i}{r_{{\rm H},i}},\\
        \bar{e}_{\rm H} &= \frac{1}{N_{\rm p}-1}\sum_{i}^{N_{\rm p}-1} \frac{e_i a_i+e_{i+1}a_{i+1}}{2r_{{\rm H},i}}.
    \end{align}
    respectively, with
    \begin{equation}
        r_{{\rm H},i} = \qty(\frac{M_i+M_{i+1}}{3M_*})^{1/3} \frac{a_i+a_{i+1}}{2}.
    \end{equation}
    \item \textbf{Standard deviation of mass and semi-major axis}\\
    To quantify the final variation of the planetary mass and the semi-major axis, we calculate the normalized standard deviation for each run, that is, 
    \begin{equation}
        \frac{\sigma_M}{\bar{M}} =\frac{1}{\bar{M}} \sqrt{\frac{1}{N_{\rm p}}\sum_i^{N_{\rm p}}(M_i-\bar{M})^2}
    \end{equation}
    and similarly for $a$.
    \item \textbf{Masses and semi-major axes of massive planets}\\
    To see the characteristics of the dominant planets in each system, we compare the masses and semi-major axes of the most and the second-most massive planets, averaging over 20 runs.
\end{enumerate}


\subsection{Results for the reference model}

\begin{figure*}
    \centering
    \includegraphics[width=2\columnwidth]{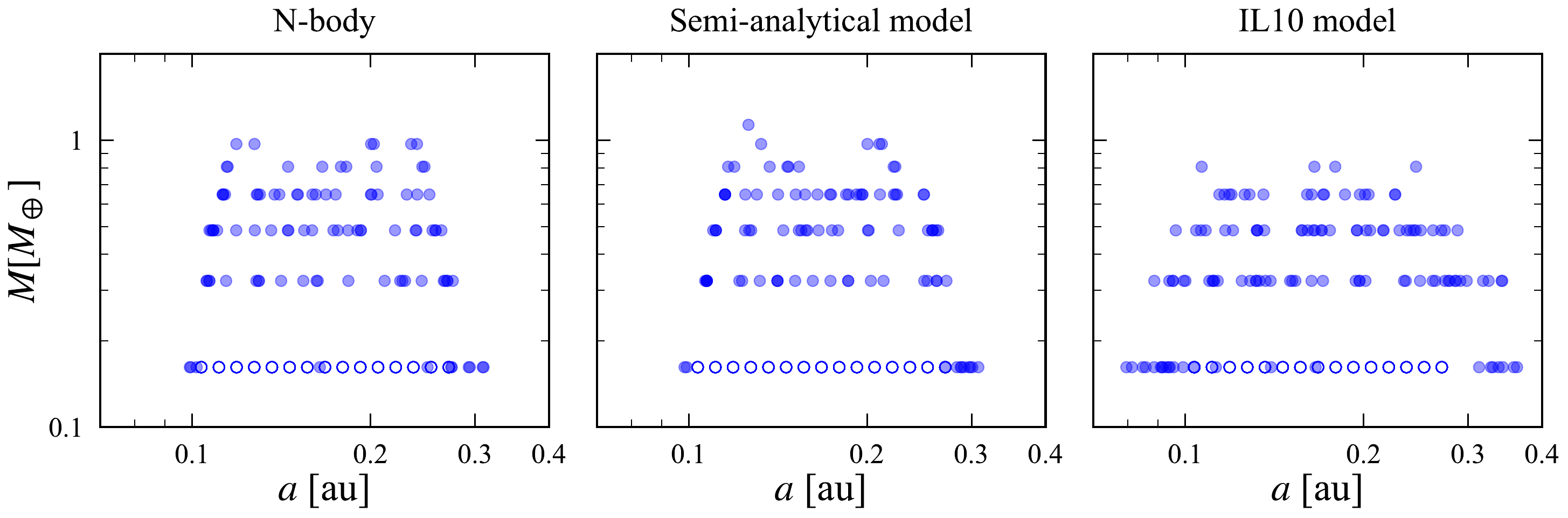}
    \caption{
    The distribution of the final semi-major axis and planetary mass for all the 20 runs. The results with the reference model (model S0) are shown.
    The left, middle, and right panels show the results of the $N$-body simulations, our semi-analytical model, and the IL10 model.
    The open circles show the initial distributions.
    }
    \label{fig:Ma_final_aM02}
\end{figure*}

Here, we compare the results of our semi-analytical model with those of $N$-body simulations for the reference model (Model S0), in which the initial orbital radii are set to 0.1--0.3~au (see Table~\ref{tab:init_cond}).
Figure~\ref{fig:Ma_final_aM02} shows the final distribution of the semi-major axis and mass for the 20 runs.
The result obtained by the same model as \cite{Ida+Lin2010} (hereafter, ``IL10 model'') is also plotted.
Our model reproduces the results of the $N$-body simulations well, especially in terms of semi-major axis spread and typical planetary masses. On the other hand, the IL10 model has a more extended semi-major axis distribution and smaller planetary masses overall.
In particular, many planets do not experience giant impacts and remain at their initial mass.

In the IL10 model, orbital crossing planets are always assumed to be scattered first, and their eccentricities are excited to $\sim e_{\rm esc}$.
Then, if their epicycle amplitudes overlap with those of some other planets, some of them collide and merge.
This picture is valid in the region of $\sim 1$~au, where the collision probability is low.
Also, because of the large value of $e_{\rm esc}$, the scattered planet collides with other planet in most cases.
However, in the closer-in region, their model significantly underestimates the collision rate because, due to the small value of $e_{\rm esc}$, the scattered planet often does not overlap its orbit with other planet.
Thus, the IL10 model predicts fewer collisions and a larger number of scattered planets, which makes the typical planetary mass smaller and the semi-major axis distribution wider.

In reality, the $N$-body simulations show that, in the close-in regions, the eccentricities of neighbouring planets are excited until their epicycle amplitudes overlap due to multi-planet interactions, and most planets collide with each other without being scattered.
By introducing the collision probability $p_{\rm col}$ and always exciting the eccentricities until the adjacent epicycle amplitudes overlap (not limited by $e_{\rm esc}$), our model successfully predicts the collision rate and reproduces the features of the results of $N$-body simulations.

In the following sections, we compare the statistics of the results of our model with those of the $N$-body simulations for various initial conditions.

\subsection{Dependence on the orbital radius}

\begin{figure}
    \centering
    \includegraphics[width=\columnwidth]{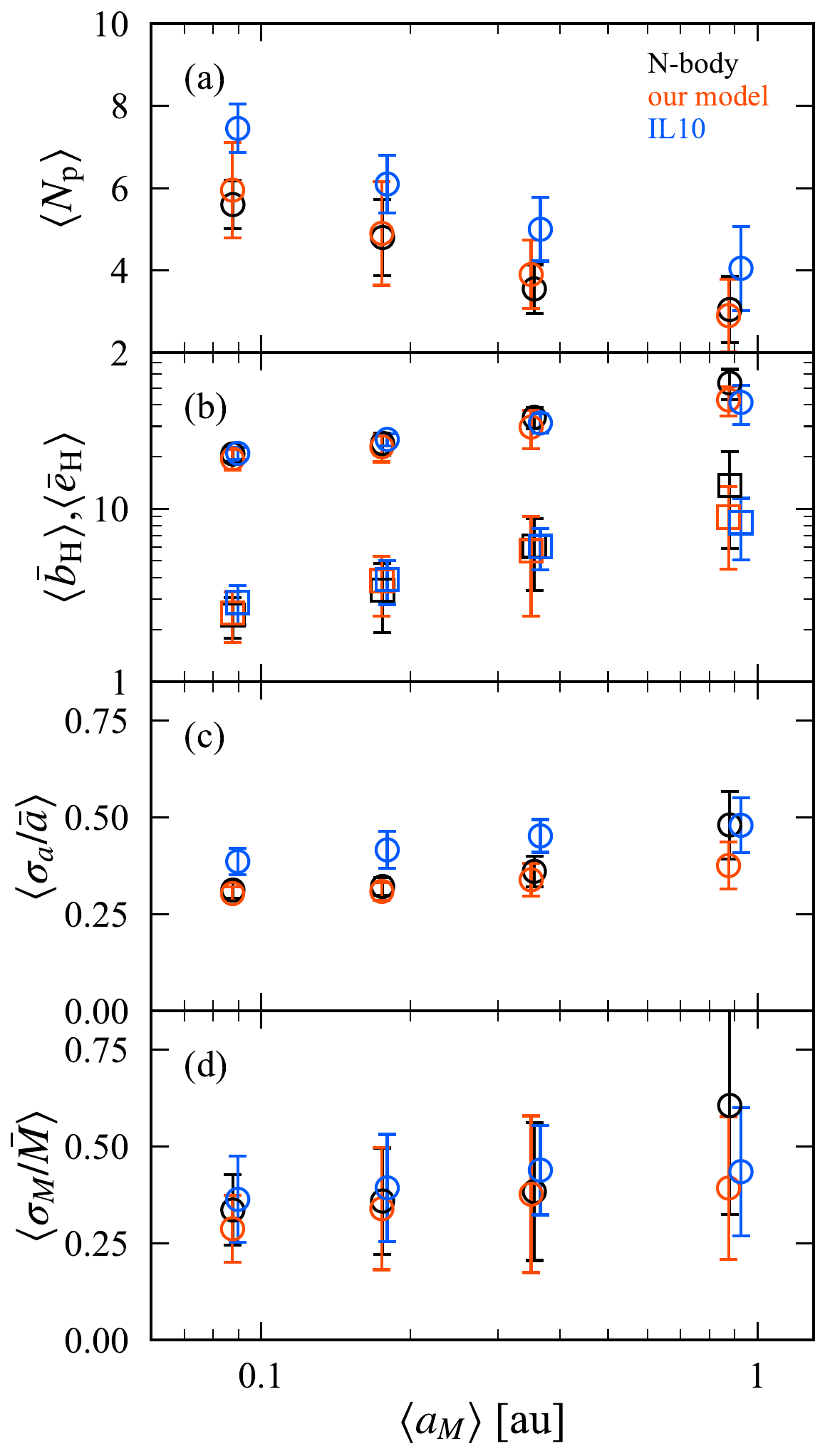}
    \caption{Comparison between the results of the $N$-body simulations (black symbols), our semi-analytical model (red symbols), and model by \cite{Ida+Lin2010}
    (blue symbols) regarding some statistical properties of the final planet distribution for Models S0 and R1--R3. The horizontal axes are the center of mass of the initial protoplanets. Each panel shows the 20-runs mean (symbols) and the standard deviation (error bars) of (a) the final number of planets $N_{\rm p}$, (b) the mean orbital separation $\bar{b}_{\rm H}$ (circles) and the mean Hill-scaled eccentricity $\bar{e}_{\rm H}$ (squares), (c) the normalized deviation of semi-major axis $\sigma_a/\bar{a}$, (d) the normalized deviation of planetary mass $\sigma_M/\bar{M}$.}
    \label{fig:all_stats_aM}
\end{figure}

Here, we show the results for different orbital radii (Model S0 and R1--R3).
Figure~\ref{fig:all_stats_aM} shows some statistical features of the final distributions in these models.
Overall, we find that our model reproduces well the results of $N$-body simulations in terms of the final number of planets, the orbital separations, the eccentricities, and deviations of mass and semi-major axis.
In contrast, the previous IL10 model tends to overestimate the number of planets and the deviation of the semi-major axis, especially in the closer-in regions.

$N$-body simulations show that the farther a planetary system is from its host star, the lower the final number of planets and, instead, the greater the orbital separation and the dispersion of the semi-major axis and mass. This is due to the chaotic evolution of orbits due to close scatterings that frequently occur at the large semi-major axis. The collision probabilities used in our model capture this trend.
In the case of the orbital crossing of two equal-mass bodies, the relative eccentricity during close encounters is calculated from Eq.~\eqref{eq:e_cross} as
\begin{equation}
    e_{ij} = \sqrt{e_{i}^2+e_{j}^2} = \frac{\sqrt{2}b_{ij}}{a_i+a_j}.
\end{equation}
Then
\begin{equation}
    \frac{e_{ij}^2}{e_{\rm esc}^2}
    =  0.07\qty(\frac{b_{\rm H}}{10})^2
        \qty(\frac{a_{ij}}{1~{\rm au}})^{-1}
        \qty(\frac{\rho}{3~{\rm g/cm^3}})^{-1/3}
        \qty(\frac{M_*}{M_\odot})^{1/3}.
    \label{eq:e_eesc}
\end{equation}
Thus, combined with Eqs.~\eqref{eq:lambda} and \eqref{eq:pcol}, the collision probability $p_{\rm col}$ largely decreases with increasing $a_{ij}$ for the same $b_{\rm H}$,
indicating that orbital crossing events in distant regions are more likely to result in close scatterings than collisions.
The physical meaning of this is as follows
(see Kokubo et al., submitted, for a more detailed analysis):
since $e_{\rm esc}$ increases with orbital radius, 
the ratio $e_{ij}/e_{\rm esc}$ decreases with distance. This reduces the ratio of collisional to scattering cross-section in the outer regions (Eqs.~\eqref{eq:tau_scat} and \eqref{eq:tau_col}).
As a result, collisions are more predominant in the close-in region while close scatterings dominate in the distant region.  
The evaluated $p_{\rm col}$ from these simple arguments is found to reproduce the statistical features of the $N$-body simulations well.

\begin{figure}
    \centering
    \includegraphics[width=\columnwidth]{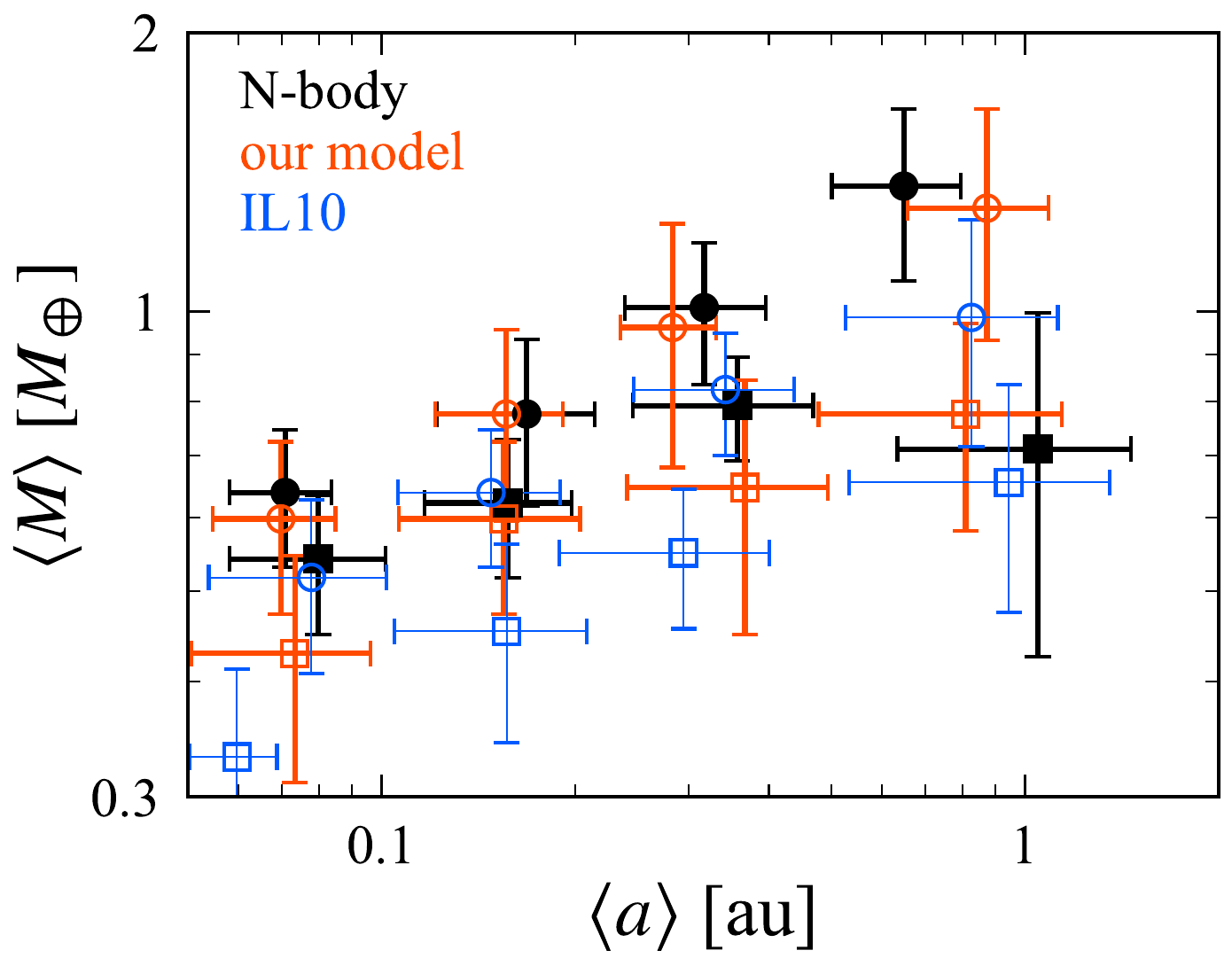}
    \caption{
    Semi-major axis $\langle a \rangle$ vs. planetary mass $\langle M \rangle$ for the most massive planet (circle) and the 2nd-most massive planet (square) in each run, averaged over 20 runs.
    The results for Models S0 and R1--R3 are shown (see Table~\ref{tab:init_cond}).
    The results of the $N$-body simulations (filled black symbols), our semi-analytical model (open red symbols), and the IL10 model (open blue symbols) are shown.
    }
    \label{fig:Mmax_aM}
\end{figure}

Figure~\ref{fig:Mmax_aM} compares the mass and semi-major axis of the most massive bodies and the second largest in each run averaged over 20 runs.
Although the previous IL10 model tends to result in less massive planets because of the lower number of collisions, our model is found to reproduce the properties of these massive planets well.

\subsection{Cases for initially equal-mass in close-in region around solar-mass stars}

\begin{figure}
    \centering
    \includegraphics[width=\columnwidth]{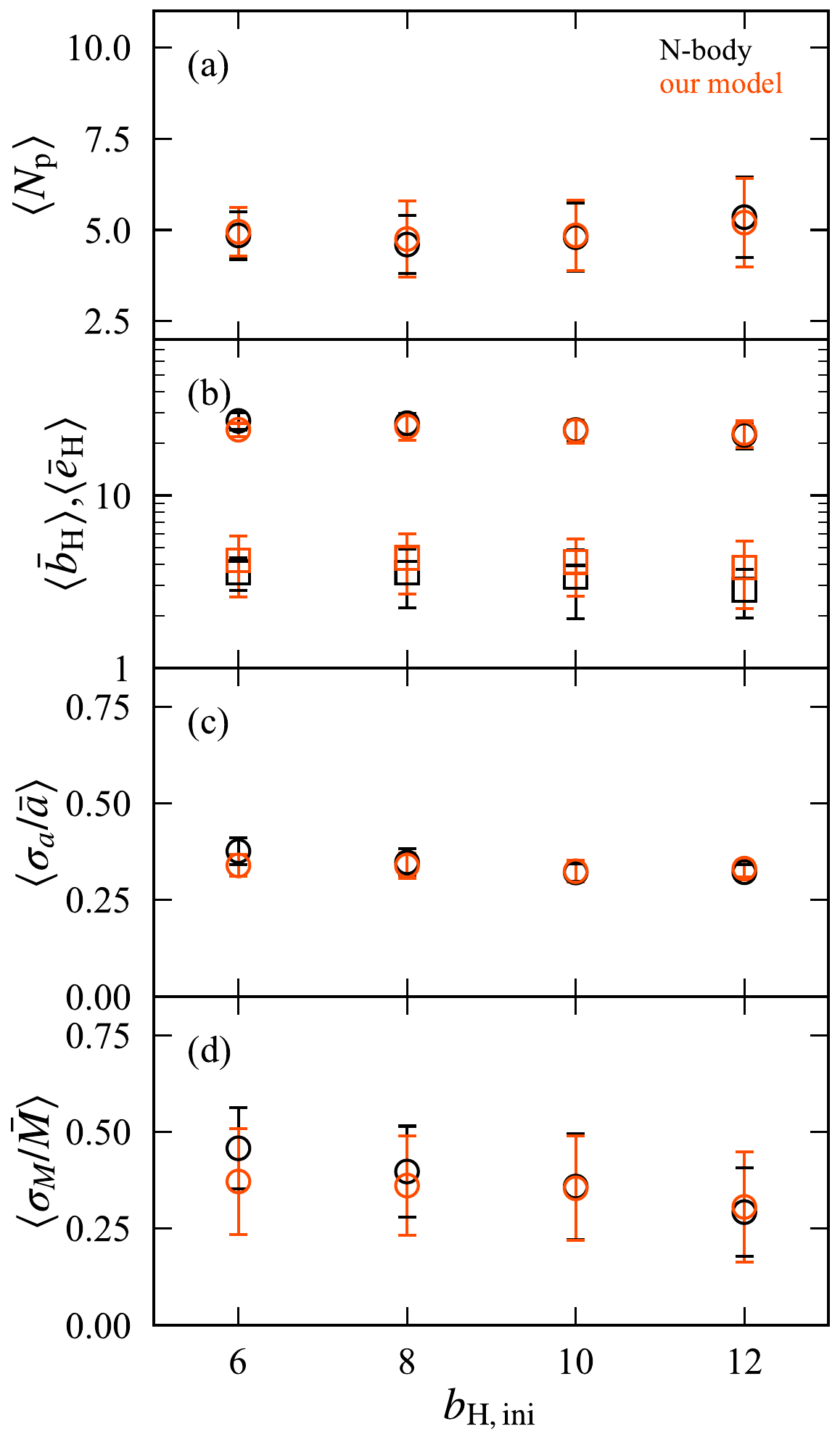}
    \caption{Same as Fig.~\ref{fig:all_stats_aM}, but for models with different initial orbital separation $b_{\rm H,ini}$ (Models S0 and B1--B3; see Table~\ref{tab:init_cond}).}
    \label{fig:all_stats_b}
\end{figure}

\begin{figure}
    \centering
    \includegraphics[width=\columnwidth]{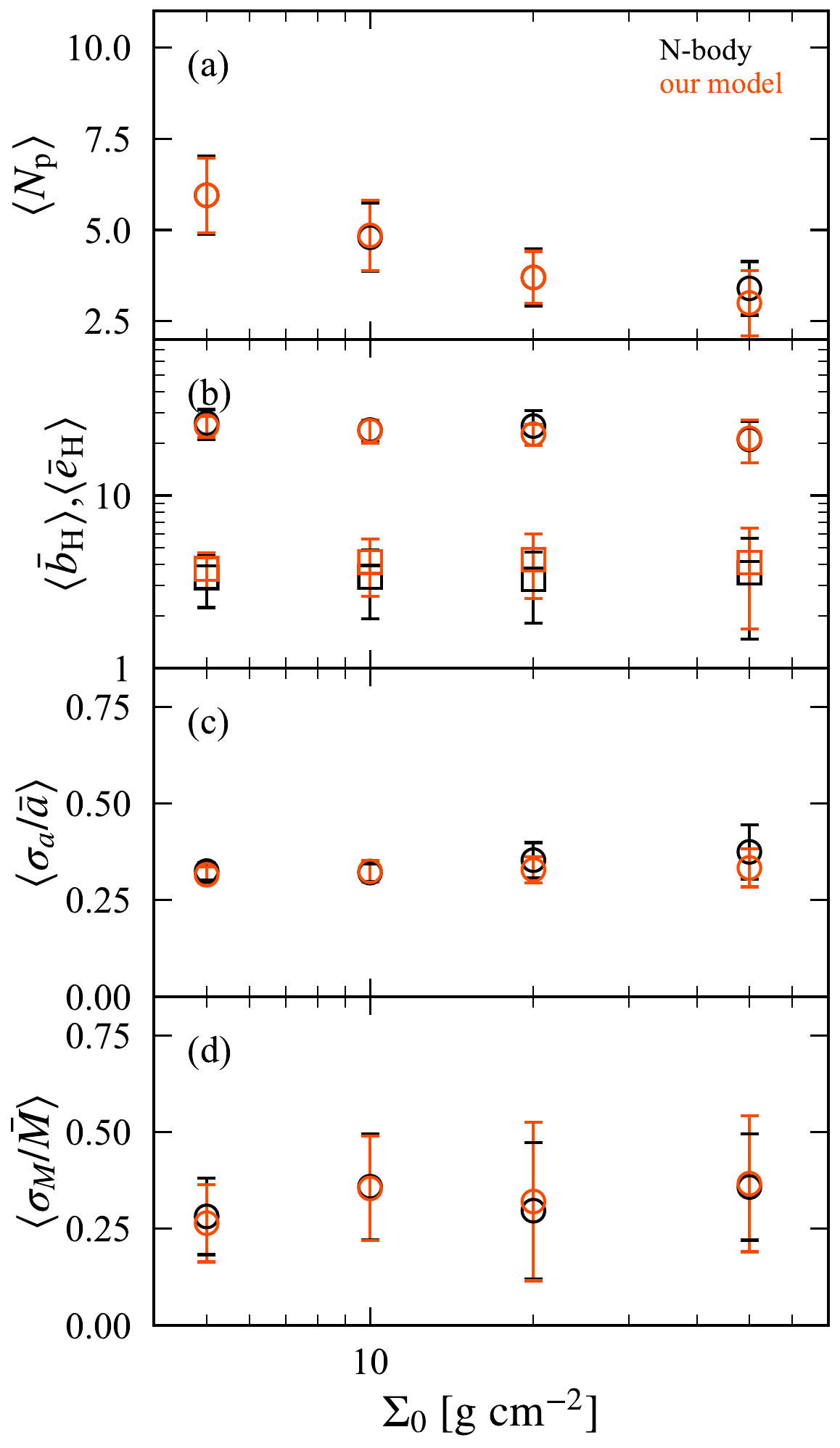}
    \caption{Same as Fig.~\ref{fig:all_stats_aM}, but for models with different solid surface density at 1~au ($\Sigma_0$)  (Models S0 and M1--M3; see Table~\ref{tab:init_cond}).}
    \label{fig:all_stats_s}
\end{figure}

Here we focus on the results with different initial orbital separations $b_{\rm H,init}$ (Models B1--B3) and different initial surface density $\Sigma_0$ (Models M1--M3), which are shown in Figs.~\ref{fig:all_stats_b} and ~\ref{fig:all_stats_s}, respectively.
All of these simulations are performed in close-in region (0.1--0.3~au) around solar-mass stars with equal-mass initial conditions.
We have found that our model can reproduce the results of the $N$-body simulations with various initial orbital separations and initial planetary masses under these equal-mass (and equal-spacing) initial conditions.

\subsection{Cases for unequal-mass initial conditions}

\begin{figure}
    \centering
    \includegraphics[width=\columnwidth]{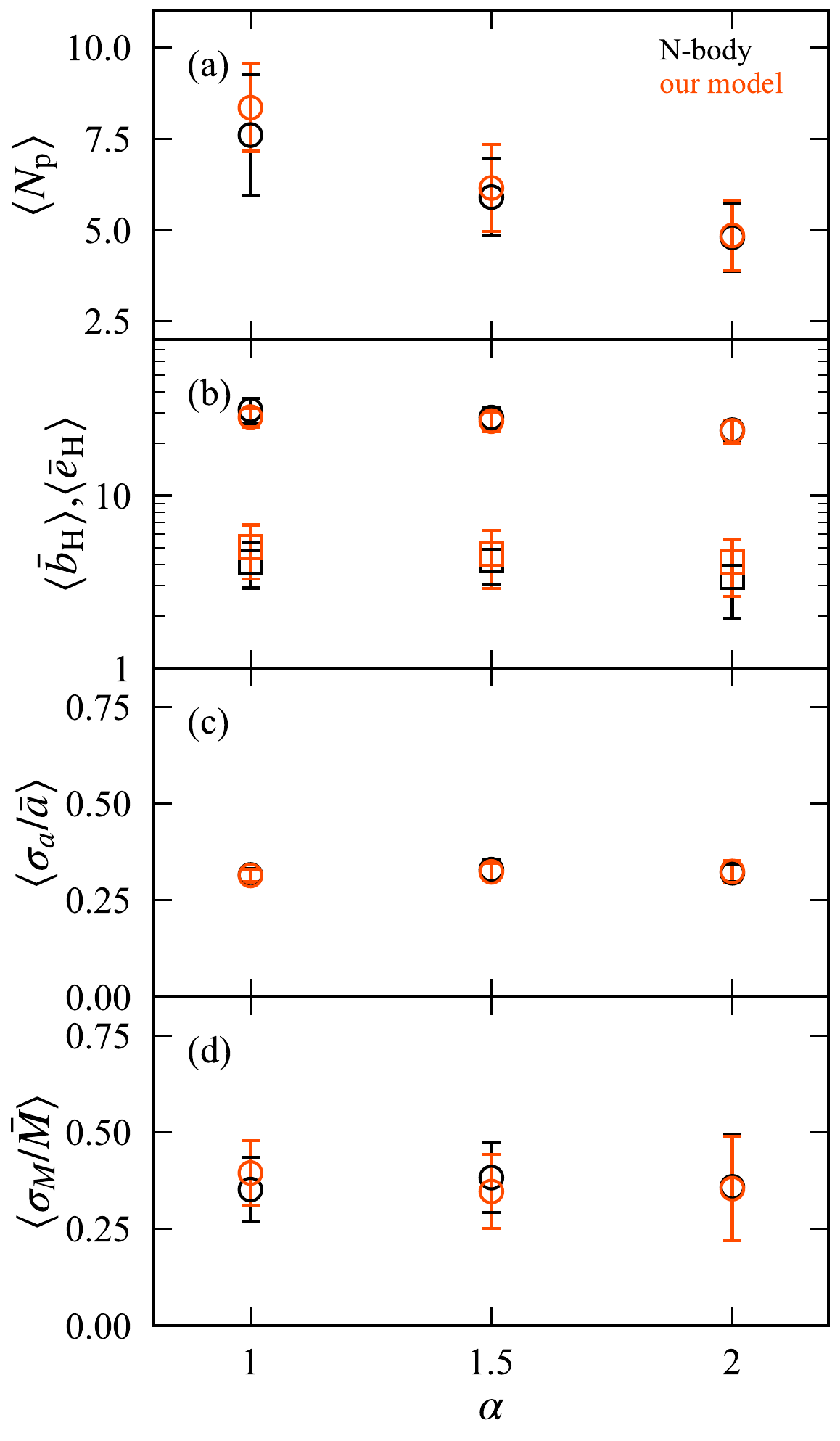}
    \caption{Same as Fig.~\ref{fig:all_stats_aM}, but for models with the different power-law index of solid surface density $\alpha$  (Models S0, A1 and A2; see Table~\ref{tab:init_cond}).
    }    
    \label{fig:all_stats_a}
\end{figure}

Here we extend our analysis to different initial solid surface density powers ($\alpha$), corresponding to cases with unequal initial planetary masses. Figure~\ref{fig:all_stats_a} presents the statistical results for Models S0, A1, and A2.
We have thus confirmed that our model is also applicable to non-equal-mass systems, which are thought to be more natural consequences of planet formation.
$N$-body simulations show that the mass dispersions $(\sigma_M/\bar{M})$ in the final state are quite similar between Models A1, A2, and S0, although the initial mass dispersion is largely different.
In particular, for Model A1 ($\alpha=1.0$), the final mass dispersion is smaller than the initial one (which is $\sim 0.5$).
This is because the 
smaller inner planets collide more frequently than the 
larger outer planets, because the eccentricities of the smaller planets are excited through gravitational perturbations from the larger bodies.
Our model successfully reproduces this feature by including the secular evolution of eccentricity.
The eccentricities of inner planets are largely excited from the initial state, leading to smaller $\tau_{\rm cross}$, and thus more collisions.

\subsection{Cases with different stellar mass}

\begin{figure}
    \centering
    \includegraphics[width=\columnwidth]{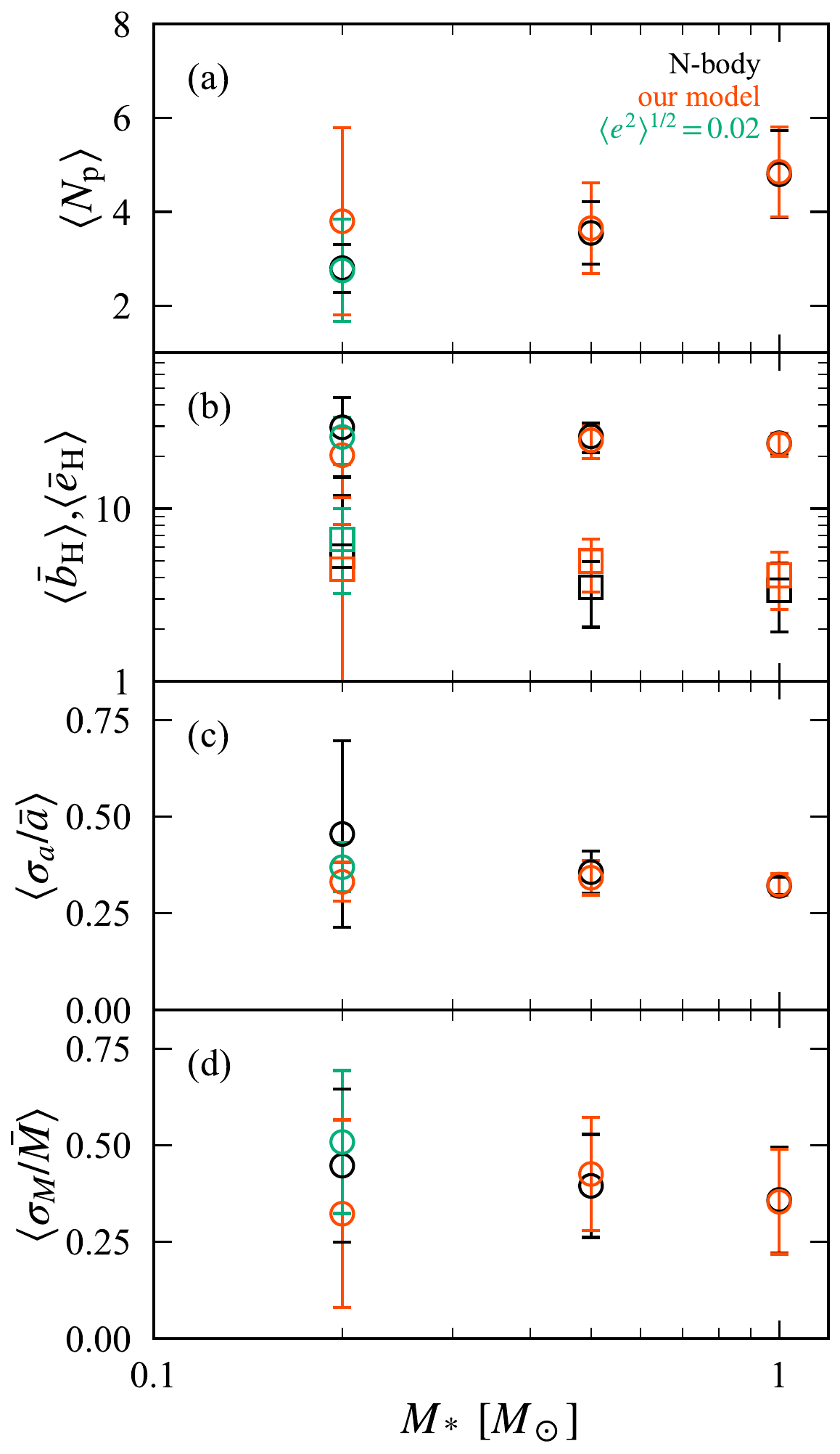}
    \caption{Same as Fig.~\ref{fig:all_stats_aM}, but for models with different central stellar mass (Models S0, S1 and S2; see Table~\ref{tab:init_cond}).
    The green points are the results when the root mean square of the initial eccentricity is set to 0.02, which is twice as large as in the other models.
    }    
    \label{fig:all_stats_Mbstar}
\end{figure}

Finally, Fig.~\ref{fig:all_stats_Mbstar} compares the results for planets around stars of different masses (Models S0, S1, and S2).
More scattering events are known to occur around lower-mass stars even in close-in regions~\citep{Matsumoto+2020, Hoshino+Kokubo2023}, leading to slightly higher values of $\langle \sigma_M/\bar{M} \rangle$ and $\langle \sigma_a/\bar{a} \rangle$ for lower-mass stars.
In our model, however, these dispersions are predicted to be smaller than the $N$-body results, and the final number of planets and its dispersion are larger in the model with $M_*=0.2M_\odot$ (Model S1).
We have found that some runs of the total of 20 runs in Model S1 show no collisional or scattering events in the integration time.
This is because the initial orbital separation $\delta$ calculated from Eq.~\eqref{eq:delta_mean} (with $\delta_{ij}$ from Eq.~\eqref{eq:delta_ij}) is, by chance, quite close to $\delta_{\rm ov}$, at which $\tau_{\rm cross}$ goes to infinity.
Whether the initial value of $\delta$ is lower than $\delta_{\rm ov}$ depends on the initial eccentricities of the planets, which are different in different runs.
Indeed, we have found that increasing the initial eccentricities in Model S1 just by a factor of two solves the discrepancy from $N$-body results, as shown in the green points in Fig.~\ref{fig:all_stats_Mbstar}.
Thus, the discrepancy would originate from our tentative way to include eccentricity to the formula of $\tau_{\rm cross}$ by \cite{Petit+2020}.
However, please note that this could be a problem only in some limited cases, such as where $\delta$ in the initial state is quite close to $\delta_{\rm ov}$.
Once close scattering or collisional events occur, our model can reproduce the $N$-body results also around lower-mass stars.
Especially, our model well-reproduces the feature that more scattering events occur around lower-mass stars because $p_{\rm col}$ is smaller for smaller $M_*$, as indicated by Eqs.\eqref{eq:lambda} and \eqref{eq:e_eesc}.

\section{Discussion} \label{sec:discussion}

\subsection{Timescale of collisional evolution}
\label{sec:timescale_of_evolution}
\begin{figure}
    \centering
    \includegraphics[width=\columnwidth]{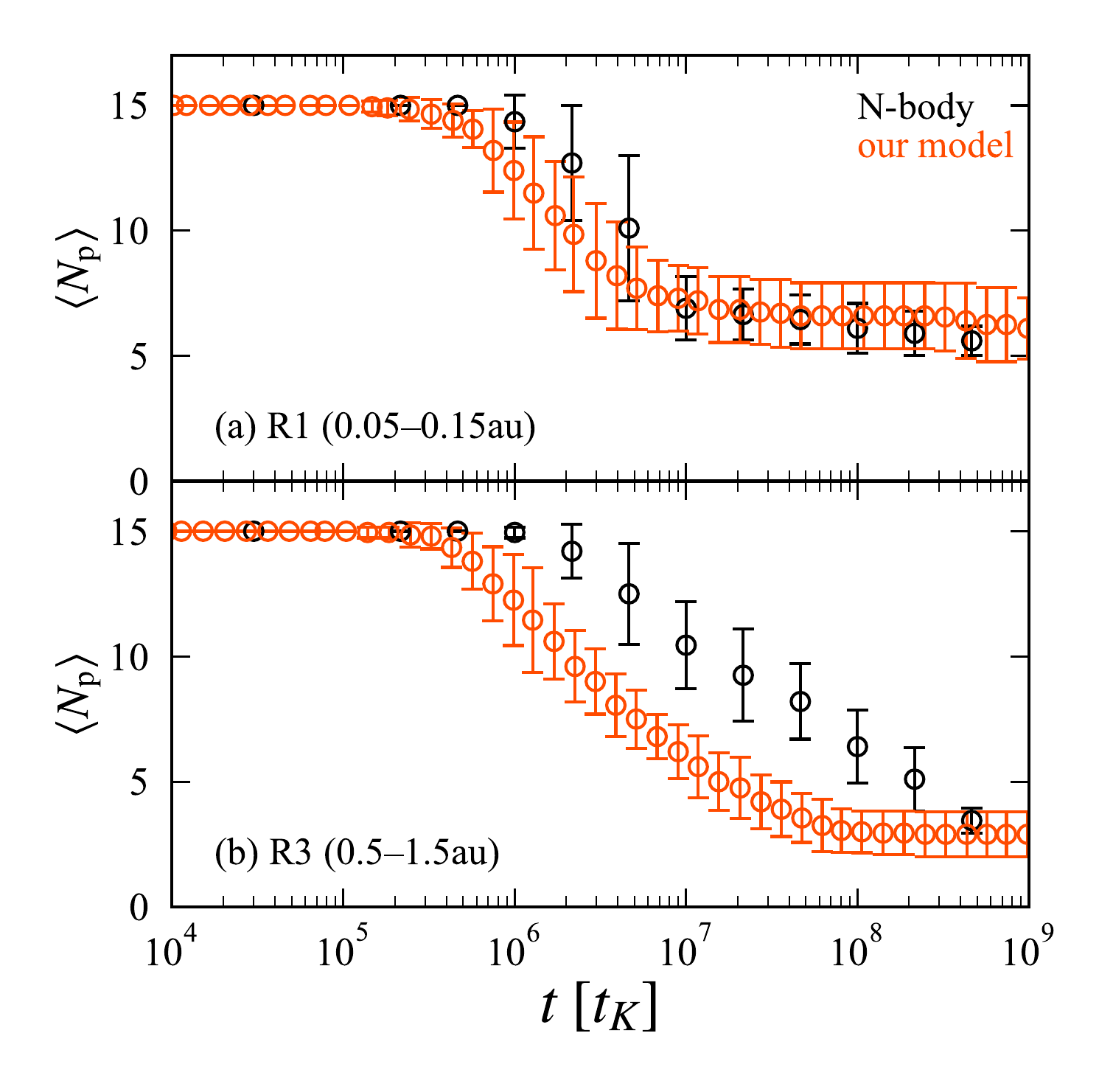}
    \caption{Evolution of the number of planets in one system averaged over 20 runs, $\langle N_{\rm p} \rangle$.
    The horizontal axes are the time normalized by the Kepler period of the innermost planet in each model.
    (a) model R1 (0.05--0.15au) and (b) model R3 (0.5--1.5au).}
    \label{fig:Nevol_aM}
\end{figure}

While we have focused on the final-state distribution resulting from collisional evolution, the evolution timescale is also an important factor in determining the properties of formed planets, particularly because orbital instability coincides with disc gas dissipation.
In cases where the growth timescale through giant impacts is shorter than the disc dissipation timescale, planets can accumulate disc gas along with mass growth, because primordial atmospheric mass is a strong function of planetary mass~\citep[e.g.,][]{Ikoma+Genda2006}. Additionally, mass growth in nebular gas can hinder subsequent atmospheric escape due to photo-evaporation, whose rate depends on planetary gravity.
Therefore, whether or not giant collisions occur in disc gas can greatly affect the final atmospheric mass and, consequently, the planetary radius.

Figure~\ref{fig:Nevol_aM} compares the time evolution of $\langle N_{\rm p}\rangle$ between our model and $N$-body simulations in different orbital radii. Notably, our model tends to underestimate the evolution timescale, particularly at larger orbital distances.
This discrepancy comes mainly from our definition of $\delta_{ij}$ (Eq.~\eqref{eq:delta_ij}), where the closest distance between two orbits (i.e., the difference between the apocenter of the inner planet and the pericenter of the outer planet) is used to represent the separation of eccentric orbits.
In fact, replacing orbital separation with the closest distance can underestimate $\tau_{\rm cross}$, depending on eccentricity and separation.
The distance between the two planets is actually equal to the closest distance only when $\Delta \varpi=\pi$, and thus the effective distance will be larger because the longitudes of the perihelion of the two planets change with time. However, the definition of appropriate separation in the case of eccentric orbits is not yet well understood.

The underestimation of $\tau_{\rm cross}$ is more significant for more distant planets since close scatterings result in larger eccentricities. 
In addition, inclinations are also excited for distant planets, and such planets are known to take a longer time from the first close encounter to the collision than in the coplanar case~\citep[e.g.,][]{Matsumoto+Kokubo2017, Rice+2018}. However, a systematic understanding of the extent to which the timescale can be longer in comparison to coplanar cases is currently lacking.

Due to these factors, our model tends to underestimate the evolution timescale, especially in scenarios with numerous close scattering events. Similar discrepancies are evident in models with small initial separation (Model B1) and smaller stellar mass (Models S1 and S2). Addressing these discrepancies requires extensive studies to establish the expression of $\tau_{\rm cross}$ for eccentric cases.



\subsection{Remaining problems}
In this study, we have considered the collisional evolution of non-resonant systems. However, taking into account the planet formation process, it is thought that multiple planets form a resonant chain around the inner edge of the disc during the formation stage because of inward migration. It is known that the timescale for the onset of orbital instability in such systems is significantly different from that of non-resonant systems. In particular, chains of planets below a certain critical number are known to be stable in the long term~\citep{Matsumoto+2012,Pichierri+Morbidelli2020, Goldberg+2022}, which is important for the understanding of currently observed exoplanetary multiplanet systems.

In addition, this study focuses on the evolution of systems consisting of planets up to several Earth masses, and it is not clear whether our model is applicable to systems with more massive planets like gas giants. 
When applying our model to such systems, it is necessary to consider the ejection of planets from the system due to scattering. 
In particular, it is known that when there are three or more gas giants, at least one planet tends to be ejected from the system as a result of orbital instability, which also has a significant impact on the orbits of the other planets~\citep{Marzari+Weidenschilling2002,Nagasawa+2008}.
Since similar ejections can also occur in super-Earths around M-type stars~\citep{Matsumoto+2020}, the appropriate treatment of such events is important for explaining the distribution of exoplanets. 
A semi-analytical model that takes into account the scattering of multiple gas giants has already been developed by \citep{Ida+2013}; however, their model tends to overestimate the number of gas giants remaining in the system due to the underestimation of the number of scattering events. 
This is due to the lack of effects of secular perturbations between gas giants. 
Our semi-analytical model with secular evolution has the potential to solve the problem. 

The dynamical evolution of close-in planets can also be affected by general relativity (GR) effects or tidal forces, both of which are not included in our $N$-body simulations or the semi-analytical model. GR is known to affect the secular evolution of the eccentricity of close-in planets, reducing the oscillation amplitude of eccentricity~\citep{Marzari+Nagasawa2020}. Although this might impact system stability, the quantitative effect of GR on the dynamical evolution of close-in planets is not well understood.

The effects of tidal forces on the simulations in this paper would be limited given the ranges of planetary mass and semi-major axis we have explored~\citep[e.g.,][]{Jackson+2009,Bolmont+Mathis2016}. However, tidal forces become crucial for planets with orbits even closer to the star ($\sim$0.05 au). Therefore, to extend the applicability of our model, the inclusion of GR and tidal effects will be an important future step.

Finally, our model assumes all collisional events result in perfect accretion. To check the validity of this assumption, we calculated the ratio between the impact velocity ($v_{\rm imp}$) and the escape velocity ($v_{\rm esc}$) for all collisional events in the $N$-body simulations in this study. We found that more than 90\% of collisional events occur with $v_{\rm imp}/v_{\rm esc} < 1.5$, which matches the condition for efficient accretion suggested by previous studies~\citep[e.g.,][]{Agnor+Asphaug2004,Genda+2015}. 
In addition, although some collisional events would result in hit-and-run depending on the impact velocity and angle, it is known that these events have little impact on the final properties of planetary systems~\citep{Kokubo2010}.
Therefore, collisional fragmentation or hit-and-run events would hardly affect the results of this study. 
However, including multiple outcomes of giant collisions will enhance the applicability of our model. 

\section{Conclusions}\label{sec:conclusion}
We have developed a new semi-analytical model to follow the collisional evolution of a compact multiplanet system and compared the results from our model with those from $N$-body simulations.
By introducing an analytic expression for collisional probability, which describes the probability that an orbit-crossing planet pair causes a collision event after repeated close encounters, our model successfully reproduces many of the statistical properties obtained from the $N$-body simulations with various initial conditions.
The major results are summarised as follows.

\begin{enumerate}
    \item The semi-analytical model reproduces the number of remaining planets, the mean orbital separations, dispersions in semi-major axes, and planet masses, in the final state distribution (i.e., the state in which the system becomes stable) of planets in $N$-body simulations with the orbital radii of 0.1--1~au around solar-mass stars.    
    Specifically, our model well describes the characteristic features seen in $N$-body simulations, where giant collisions dominate for planets near the central star, and, conversely, close scattering dominates for more distant planets.
    \item We have confirmed that the model is applicable to planetary systems with various initial conditions, such as with different initial orbital separations, larger initial masses, and non-equal initial masses.
    \item Our model is also found to be valid for close-in planets around lower-mass stars.
\end{enumerate}

As the above results show, the semi-analytical model developed in this study can predict the distribution of planetary masses and orbits after collisional evolution in a relatively wide range of orbital radii, including close-in orbits, where many exoplanets are currently found. 
This model can be used to develop integrated models of planet formation that are computationally much less expensive than conventional ones. 
Such models would enable us to constrain the planet formation process through a large number of parameter studies and comparisons with observational data.

Since this study is limited to comparisons for results from idealized initial conditions, the applicability to planetary systems derived from models of planetary growth and orbital evolution in protoplanetary disks (i.e., systems with larger mass dispersions, variations in orbital spacing, and resonant planets) will be investigated in a subsequent paper.


\begin{acknowledgments}
We thank A. Petit for fruitful discussion.
This work is supported by JSPS KAKENHI Grant Nos. 18H05438, 18H05439, 22KJ0816, 22K21344, 24K00698 and 24H00017.
\end{acknowledgments}

%






\appendix
\section{Secular perturbation theory}
\label{sec:secular_theory}
Here we briefly summarize the Laplace-Lagrange secular perturbation theory~\citep[e.g.,][]{Murray+Dermott1999}, and describe how to calculate the evolution of eccentricity in our model.
Assuming that $e_i \ll 1$ and $I_i=0$, the secular term of the disturbing function for planet $i$ in an $N$-body system, expanded to the second order of $e_i$, is given by
\begin{equation}
    \mathcal{R}_i = n_i a_i^2 \qty[
    \frac{1}{2}A_{ii}(h_i^2 + k_i^2) + 
    \sum_{j=1, j\neq i}^N A_{ij}(h_i h_j + k_i k_j)
    ].
\end{equation}
Here $\mathcal{R}_i$ is expressed as a function of the eccentricity vector:
\begin{equation}
    h_i = e_i \sin \varpi_i, \quad
    k_i = e_i \cos \varpi_i,
\end{equation}
with $\varpi_i$ being the longitude of pericenter.
The coefficients are given by
\begin{align}
    A_{ii} &= n_i \frac{1}{4}\sum_{k=j,j\neq i}^N \frac{M_j}{M_* + M_i}\alpha'_{ij}\bar{\alpha'}_{ij} b_{3/2}^{(1)}(\alpha'_{ij}), \label{eq:Ajj_Nbody}
    \\
    A_{ij} &= -n_i \frac{1}{4}\frac{M_j}{M_* + M_i}\alpha'_{ij}\bar{\alpha'}_{ij} b_{3/2}^{(2)}(\alpha'_{ij})    
    \label{eq:Ajk_Nbody}
\end{align}
where $n_i$ is the mean motion, and $\alpha'_{ij}, \bar{\alpha'}_{ij}$ are the ratios of semi-major axes:
\begin{align}
    \alpha'_{ij} &= 
    \begin{cases}
        a_i/a_j  &\qif a_i < a_j , \\
        a_j/a_i  &\qif a_i > a_j   
    \end{cases}\\
    \bar{\alpha'}_{ij} &= 
    \begin{cases}
        \alpha'_{ij} &\qif a_i < a_j , \\
        1           &\qif a_i > a_j.
    \end{cases}
\end{align}
Finally, the Laplace coefficient $b_{s}^{(m)}(a)$ is defined as
\begin{equation}
    b_{s}^{(m)} (a ) = 
    \frac{2}{\pi} \int_0^{\pi}
    \frac{\cos m\varphi}{(1+a^2-2a\cos \varphi)^s}
    \dd{\varphi}.
\end{equation}
The calculations of the Laplace coefficients are performed by using a Python package called \verb|PyLaplace|.

Then, the evolution of the eccentricity vector is expressed as
\begin{align}
    \dot{h}_i = +\frac{1}{n_i a_i^2}\pdv{\mathcal{R}_i}{k_i}, \quad
    \dot{k}_i = -\frac{1}{n_i a_i^2}\pdv{\mathcal{R}_i}{h_i},
\end{align}
which can be further reduced to
\begin{align}
    \dot{h}_i = \sum_{j=1}^N A_{ij}k_j, \quad
    \dot{k}_i = -\sum_{j=1}^N A_{ij}h_j    
\end{align}
Thus, using the eigenvalues $\bm{g} = (g_1, g_2, \dots, g_N)$ of the $N\times N$ matrix $\bm{A}=\{A_{ij}\}$ and the matrix $\bm{e} = (\bm{e}_1, \bm{e}_2, \dots , \bm{e}_N)$ storing the corresponding eigenvectors (where the element $e_{ij}$ represents the $i$-th component of the eigenvector $\bm{e}_j$), the solution is
\begin{equation}
    h_i = \sum_{j=1}^N e_{ij}\sin (g_j t + \beta_j), 
    \quad
    k_i = \sum_{j=1}^N e_{ij}\cos (g_j t + \beta_j).
\end{equation}
Here the eigenvectors $\bm{e}_j$ and the initial phase angle $\beta_j$ for each planet are determined from certain boundary conditions.
Then, the eccentricity of each planet is evaluated as
\begin{equation}
    e_i(t) = \sqrt{h_i(t)^2 + k_i(t)^2}.
\end{equation}

\section{Orbital instability timescale by Petit et al. (2020)}
\label{sec:tcross_by_Petit}
Here we describe the explicit form of the analytical formula of $\tau_{\rm cross}$ given by \cite{Petit+2020}. Planetary orbits in a system with three or more planets are supposed to deviate from their initial state because of perturbations from neighbouring planets via three-body mean motion resonances. When a planetary orbit crosses the location of a two-body resonance, the planet experiences strong perturbations, resulting in significant changes in its orbit over a short timescale.
Thus, the diffusion (in terms of the semi-major axis) via three-body resonances determines the orbital crossing timescale.

Based on their analysis, \cite{Petit+2020} derived the analytical formula for $\tau_{\rm cross}$ in a system with three circular-orbit planets (denoted as planets 1, 2, and 3, with $a_1 < a_2 < a_3$) as
\begin{align}
    \log(\frac{\tau_{\rm cross}}{P_1})
      &= -\log(\frac{32\sqrt{19}M\sqrt{\eta (1-\eta)}}{3\sqrt{\pi}}) 
        + \log(\frac{\delta^6}{\delta_{\rm ov}^6}
      \frac{1}{1-(\delta/\delta_{\rm ov})^4}) 
      + \sqrt{-\ln\qty[1-\qty(\frac{\delta}{\delta_{\rm ov}})^4]}
      \label{eq:tau_cross_Petit}
\end{align}
with 
\begin{align}
    M &= \frac{\sqrt{M_1M_3+M_2M_3\eta^2\alpha_{12}^{-2}
    +M_1M_2\alpha_{23}^2(1-\eta)^2}}{M_*}, \\
    \delta &= \frac{\delta_{12}\delta_{23}}{\delta_{12}+\delta_{23}},\label{eq:delta_mean} \\
    \eta &= \frac{\nu_{12}(1-\nu_{23})}{1-\nu_{12}\nu_{23}}, \\
    \delta_{\rm ov} &= (6.55KM)^{1/4}\qty[\eta(1-\eta)]^{3/8}.
    \label{eq:delta_ov}
\end{align}
Here $\alpha_{ij}=a_i/a_j$ and $\nu_{ij}=P_i/P_j$ are the ratios of the semi-major axes and the orbital periods, respectively, and $\delta_{ij}$ is the normalized orbital separation, which is defined as $\delta_{ij} = 1-\alpha_{ij}$ in \cite{Petit+2020}.
The numerical factor $K$ in Eq.~\eqref{eq:delta_ov} accounts for the influence of the number of planets in a system on the density of three-body resonances.
See main text for the prescriptions of $\delta_{ij}$ and $K$ in our semi-analytical model.
Note that Eq.~\eqref{eq:tau_cross_Petit} can only be defined at $\delta_{ij} < \delta_{\rm ov}$, where $\delta_{\rm ov}$ is a critical orbital separation beyond which the density of three-body resonances rapidly decreases, leading to system stability. Thus,
$\tau_{\rm cross}$ approaches infinity when $\delta \to \delta_{\rm ov}$.


\bibliography{refer}{}
\bibliographystyle{aasjournal}



\end{document}